\RequirePackage[2020-02-02]{latexrelease}
\documentclass[twocolumn,preprintnumbers,amsmath,amssymb,pra]{revtex4}

\usepackage{graphicx,amsmath}
\usepackage{dcolumn}
\usepackage{bm}
\usepackage{amssymb}
\usepackage{epstopdf}
\usepackage{color}
\graphicspath{{Figures/}}

\begin{document}

 \title{Decay of transmon qubit in a broadband one-dimensional
 cavity}

\begin{abstract}

We investigate the decay dynamics of a three-level artificial
atom, a superconducting transmon qubit, weakly coupled to a
continuum of modes in a broadband, one-dimensional cavity. Using
the resolvent formalism, we derive analytical expressions for the
resonance frequency shifts and widths, which are then evaluated
numerically for a Gaussian density of states. We identify two
distinct dynamical regimes, differentiated by the ratio of the
qubit's coupling strength to the continuum bandwidth. When this
ratio is much less than one, the system exhibits a Markovian
regime in which the resonance width is practically independent of
energy within the continuum band. As the ratio increases, the
system transitions to a non-Markovian regime where the resonance
width becomes strongly energy-dependent. In this regime, the qubit
interacts with the continuum faster than the continuum can erase
the information from the qubit's past. Furthermore, we demonstrate
that the coupling between the transmon's second level and its
ground state significantly influences the decay dynamics of the
third level. The interaction between these two levels opens a fast
two-photon decay channel, which broadens the transmon's second
level.

\end{abstract}



\author{Ya. S. Greenberg}\email{yakovgreenberg(c)yahoo.com}
\affiliation{Department of Applied and Theoretical Physics,
Novosibirsk State Technical University, Novosibirsk 630073,
Russia}

\author{A. A. Shtygashev} \affiliation{Department of Applied and Theoretical Physics,
Novosibirsk State Technical University, Novosibirsk 630073, Russia}

\author{O. V. Kibis} \affiliation{Department of Applied and Theoretical Physics,
Novosibirsk State Technical University, Novosibirsk 630073, Russia}

\maketitle

\section{Introduction}

In recent years, the superconducting transmon qubit has become one
of the most promising platforms for the realization of scalable
quantum processors \cite{Ach2024,Ach2023,Arute2019,Jurc2021}. The
transmon is fabricated using thin-film technology and operates as
a three-level artificial atom. This design is very robust against
bias-voltage noise \cite{Koch2007} and can achieve an extremely
long coherence time \cite{Wang2022}.

Recent developments in quantum technology have enabled new
experimental implementations of transmon-photon interactions.
These implementations are based on the strong coupling achievable
between microwave photons and qubits in an open coplanar waveguide
\cite{Astaf2010, Sult2025, Gasp2017, Lu2021, Gunin2025}.

The efficiency of photon manipulation in a waveguide-qubit setup
depends on the qubit decoherence rate. Standard measurement
procedures based on dispersive qubit readout, which are commonly
used in a cavity, cannot be directly applied to an open waveguide
due to the lack of a resonator.

The measurement of the relaxation and decoherence rates of a
superconducting transmon qubit in an open waveguide requires the
application of a two-pulse technique that excites both transmon
levels \cite{Sult2025}. On the other hand, intense-pulse
measurements in a resonator-free setting allow for the
investigation of the effects arising from strong coupling between
a transmon and the continuum of waveguide modes \cite{Chang2007}.

In this paper, we investigate the decay dynamics of a three-level
artificial atom, a superconducting transmon qubit, interacting
with a continuum of modes in a waveguide. The waveguide is treated
as a broadband, one-dimensional cavity with a low quality factor.

We focus on the population dynamics of the transmon's third level
by calculating its survival probability. This is the probability
that the system, initially excited to this level, is still found
there at time $t$.

Using the resolvent and projector operator formalism, we obtain
analytical expressions for the resonance shifts and widths for
transmon's third level. These values are then calculated
numerically for a Gaussian density of states. We also show that
the decay dynamics of the transmon differ drastically from those
of a two-level system, as the coupling between the transmon's
second level and its ground state significantly broadens
transmon's third level.

We identify two distinct regimes, differentiated by the ratio of
the qubit's coupling strength to the continuum bandwidth. When
this ratio is much less than one, the system exhibits a Markovian
(weak-coupling) regime, in which the resonance width is
practically independent of energy within the continuum band.

As the ratio increases, the system transitions to a regime where
the resonance width becomes strongly energy-dependent. This is the
non-Markovian (strong-coupling) regime, where the qubit interacts
with the continuum faster than the continuum can erase the
information from the qubit's past.

It is worth noting that our definitions of weak- and
strong-coupling regimes should not be confused with conventional
definitions generally adopted in cavity QED. In cavity QED, these
regimes are defined by the ratio of the transmon's coupling
strength to its excitation frequencies. Throughout this paper, we
remain in the conventional weak-coupling regime, where the
transmon's coupling strength is much less than its excitation
frequencies.

The paper is organized as follows. The model is described in
Section II, where the interaction between the transmon qubit and
the continuum is formulated as a multimode Jaynes-Cummings
Hamiltonian. In Section III, we formulate the projection operator
technique for our problem.  In Section IV we calculate the matrix
element for the shift operator of the third transmon level. The
principal difference between the decay of a transmon and that of a
two-level system is studied in Section V using the Dyson equation
for the resolvent. Analytical expressions for resonances and
widths of the third transmon level are calculated in Section VI.
 In Section VII, we transform the expressions for the frequency shifts
and resonance widths into the continuum limit. Section VIII
studies the coupling between the transmon qubit and continuum
modes for a Gaussian density of states. The spectral function
obtained in Section VIII allows us to numerically calculate the
resonance frequencies and their widths in Section IX.

In Appendix B the Green's function method is applied to obtain
general expressions for the probability amplitudes of the transmon
levels. The influence of the higher transmon levels on the decay
of transmon qubit is considered in Appendix B.

\section{The model}

The transmon is an anharmonic oscillator with nearest-neighbor
coupling. In this paper, we consider only the first three levels
of the transmon. In this configuration, it functions as a transmon
qubit-a three-level artificial atom. For conciseness, throughout
the paper we will frequently refer to the transmon qubit simply as
the transmon.

\begin{figure}
  \includegraphics[width=8 cm]{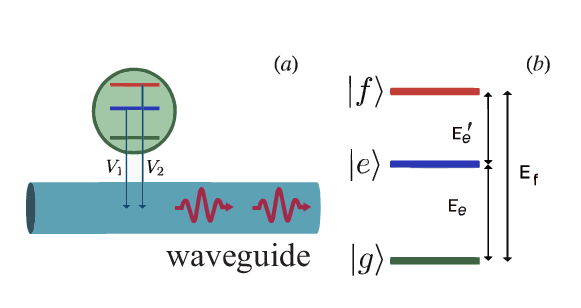}\\
  \caption{The interaction of transmon qubit with a waveguide (a)
  and its level structure (b).}\label{Fig1}
\end{figure}

We denote these levels as $|g\rangle, |e\rangle, |f\rangle$ with
the energies $E_g$, $E_e$, $E_f$. Throughout the
paper we measure energies from the ground state $|g\rangle$, setting
$E_g=0$. The level structure of transmon qubit is
shown in Fig.\ref{Fig1}.

The distinctive feature of a transmon is that the  the energy
difference between the third level $|f\rangle$ and the second
level denoted $|e\rangle$, $E_f-E_e\equiv E_e^{\prime}$, is
slightly smaller than the difference between the second level
$|e\rangle$ and the ground state $|g\rangle$. Consequently, a
transmon is characterized by its anharmonicity $\Delta
E_e=E_e-E_e^{\prime}$, which can be expressed in dimensionless
unit  $\alpha=\Delta E_e/E_e$. The transmon energies $E_e$ and
$E_f$ fall within GHz range. For practical transmons the values of
$\alpha$ typically lie within the range $0.001\div 0.01$.

In what follows we set $\hbar=1$, so that all energies are
expressed in frequency units.

We write the transmon Hamiltonian in the basis of the uncoupled
transmon states as a multimode Jaynes- Cummings model
\cite{Koch2007}:

\begin{equation}\label{1}
    H = H_0  + V_1  + V_2
\end{equation}

where

\begin{equation}\label{2}
H_0  = E_e \left| e \right\rangle \left\langle e \right| + E_f
\left| f \right\rangle \left\langle f \right| + \sum\limits_k {}
\omega _k a_k ^ +  a_k
\end{equation}

\begin{equation}\label{3}
V_1  = \sum\limits_k {} g_1 (k)\left( {\left| g \right\rangle
\left\langle e \right|a_k ^ +   + \left| e \right\rangle
\left\langle g \right|a_k } \right)
\end{equation}

\begin{equation}\label{4}
V_2  = \sum\limits_k {} g_2 (k)\left( {\left| e \right\rangle
\left\langle f \right|a_k ^ +   + \left| f \right\rangle
\left\langle e \right|a_k } \right)
\end{equation}

The interaction terms $V_1$ and $V_2$ describe the coupling of the
transmon to the  waveguide photon modes where $a_k^+$ and $a_k$
are the creation and annihilation photon operators, respectively.
Hamiltonian (\ref{1}) is written in the rotating wave
approximation (RWA). This approximation is valid if $g_1(k)\ll
E_e$ and $g_2(k)\ll E_f$; we assume these conditions hold
throughout the paper. In this regime, the transmon is considered
to be weakly coupled to the waveguide modes.

In a transmon, the nearest-neighbor coupling is realized
\cite{Koch2007}. Therefore, for the first three levels considered
here the second level $|e\rangle$ interacts with both the ground
state $|g\rangle$ and the third level $|f\rangle$, while direct
interaction between $|g\rangle$ and $|f\rangle$ is absent.
Furthermore, $g_2(k)=\sqrt{(3/2)}g_1(k)$ \cite{Koch2007}.

\section{Matrix elements of the resolvent operator}

In order to find the matrix elements of the resolvent operator
$G(z)$ we apply the projection operator formalism
\cite{Kohen2004}, a method well established for studying similar
problems \cite{Hu2023, Green2015}.

 From the space of all possible states we select those that
play an important role in the physical process under study.
Specifically, we choose the following eigenvectors of the
unperturbed Hamiltonian  $H_0$: two discrete states $|e0\rangle,
|f0\rangle$, and the continuum states $|gk\rangle, |ek\rangle$,
and $|gk_1k_2\rangle$. The states $|e0\rangle$, and $|f0\rangle$
contain zero photons. The states $|gk\rangle$,and $|ek\rangle$
contain one photon with wave vector $k$, while the state
$|gk_1k_2\rangle$ describes the transmon in its ground state with
two photons of wave vectors $k_1$ and $k_2$ in the system.
Clearly,

\begin{equation}\label{8}
\begin{gathered}
  \left| {gk} \right\rangle  = a_k^ +  \left| {g0} \right\rangle  \hfill \\
  \left| {ek} \right\rangle  = a_k^ +  \left| {e0} \right\rangle  \hfill \\
  \left| {gk_1 k_2 } \right\rangle  = \frac{1}
{{\sqrt 2 }}a_{k_1 }^ +  a_{k_2 }^ +  \left| {g0} \right\rangle  \hfill \\
  \left\langle {gk_1 k_2 |gk_3 k_4 } \right\rangle  = \frac{1}
{2}\left( {\delta _{k_1 ,k_3 } \delta _{k_2 ,k_4 }  + \delta _{k_1 ,k_4 } \delta _{k_2 ,k_3 } } \right) \hfill \\
\end{gathered}
\end{equation}

Next,we define the projection operators $P$ and $Q$

\begin{equation}\label{P}
P = \left| {e0} \right\rangle \left\langle {e0} \right| + \left|
{f0} \right\rangle \left\langle {f0} \right|
\end{equation}

\begin{equation}\label{Q}
    Q = \sum\limits_k {} \left| {gk} \right\rangle \left\langle {gk}
\right| + \sum\limits_k {} \left| {ek} \right\rangle \left\langle
{ek} \right| + \sum\limits_{k_1 ,k_2 } {} \left| {gk_1 k_2 }
\right\rangle \left\langle {gk_1 k_2 } \right|
\end{equation}

with the properties

\begin{equation}\label{7}
\begin{gathered}
  P + Q = 1;\;P^2  = P;\;Q^2  = Q;\;PQ = QP = 0; \hfill \\
  \left[ {Q,H_0 } \right] = \left[ {P,H_0 } \right] = 0 \hfill \\
\end{gathered}
\end{equation}

The operator $P$ projects onto states with no photons, while the
operator $Q$ projects onto continuum states with one or two
photons.

Using the projection operators, we obtain the following
expressions for the matrix elements $G_{ee}(z)=\langle e0|
G(z)|e0\rangle$ and $G_{ff}(z)=\langle f0|G(z)|f0\rangle$ (see
Appendix A)

\begin{equation}\label{14}
\begin{gathered}
  G_{ee} (z) = \frac{1}
{{\left( {z - E_e  - R_{ee} (z)} \right)}};\hfill \\
  G_{ff} (z) = \frac{1}
{{\left( {z - E_f  - R_{ff} (z)} \right)}} \hfill \\
\end{gathered}
\end{equation}

where the quantities  $R_{ff}=\langle f0|R(z)|f0\rangle$,
$R_{ee}=\langle e0|R(z)|e0\rangle$  are the matrix elements of the
shift operator $R(z)$, Calculating these  for the interaction
$V=V_1+V_2$ is the primary goal of our study.

It might seem from (\ref{14}) that the decay of a given transmon
level depends on its interaction with the nearest level which lies
below. In fact, as we show below, the decay of second transmon
level $|e\rangle$ depends only on its interaction $V_1$ with the
ground state $|g\rangle$, while the decay of the third level
depends both on $V_1$ and $V_2$.

The probability $U_{ff}(t)$
 for the system to remain in its initial state $|f0\rangle$
 after a time $t$ is given by:

\begin{equation}\label{15}
U_{ff} (t) = \int_{ - \infty }^{ + \infty } {dE} \;U_{ff} (E)e^{ -
iEt}
\end{equation}
where
\begin{equation}\label{16}
  U_{ff} (E) = \frac{1}
{{2\pi i}}\mathop {\lim }\limits_{\eta  \to 0_ +  } \left( {G_{ff}
(E - i\eta ) - G_{ff} (E + i\eta )} \right),
\end{equation}

\begin{equation}\label{17}
  G_{ff} (E \pm i\eta ) = \frac{1}
{{\left( {E \pm i\eta  - E_f  - R_{ff} (E \pm i\eta )} \right)}}
\end{equation}

In these equations the subscript $ff$ means the diagonal matrix
element between states $|f0\rangle$. For example, $U_{ff}\equiv
\langle f0|U(t)|f0\rangle$, etc.

Applying here the general form of the shift operator
\cite{Gold1964}

\begin{equation}\label{18}
  R_{ff} (E \pm i\eta ) = \Delta _2 (E,\eta ) \pm i\frac{{\Gamma _2 (E,\eta )}}
{2}
\end{equation}

we write the expression (\ref{17}) in the following form

\begin{equation}\label{19}
G_{ff} (E \pm i\eta ) = \frac{1} {{E \pm i\eta  - E_f  - \Delta _2
(E,\eta ) \pm i\frac{{\Gamma _2 (E,\eta )}} {2}}}
\end{equation}

Substituting (\ref{19}) into (\ref{16}) yields:

\begin{equation}\label{20}
\begin{gathered}
U_{ff} (E)\hfill\\ = \frac{1} {{2\pi i}}\mathop {\lim
}\limits_{\eta \to 0} \frac{{2i\eta  + i\Gamma _2^{} (E,\eta )}}
{{\left( {E - E_f  - \Delta _2 (E,\eta )} \right)^2  +
\frac{{\Gamma _2^2 (E,\eta )}} {4}}}
\end{gathered}
\end{equation}

The explicit expressions for $\Delta_2(E,\eta)$ and
$\Gamma_2(E,\eta)$ will be calculated below.

\section{Calculation of the matrix elements of the shift operator}

The shift operator $R(z)$ can be written in the form of infinite
series of powers of the interaction $V$ \cite{Kohen2004}:

\begin{equation}\label{Rz}
\begin{gathered}
  R(z) =  V + V\frac{Q} {{z - H_0
}}V + V\frac{Q} {{z - H_0 }}V\frac{Q}
{{z - H_0 }}V \hfill \\
   + V\frac{Q}
{{z - H_0 }}V\frac{Q} {{z - H_0 }}V\frac{Q} {{z - H_0 }}V\hfill\\
+ V\frac{Q} {{z - H_0 }}V\frac{Q} {{z - H_0 }}V\frac{Q} {{z - H_0
}}V\frac{Q}
{{z - H_0 }}V + ....... \hfill \\
\end{gathered}
\end{equation}
In what follows we use the concise form of this expression:
\begin{equation}\label{21}
R(z) = V + V\frac{Q} {{z - H_0 }}R(z)
\end{equation}

The calculation of the matrix elements of $R(z)$ involves the
calculations of the matrix elements of interaction $V=V_1+V_2$.
The interaction $V$ does not change the number of excitations.
Therefore, within the subspace subtended by projectors $P$ and $Q$
there are only three nonzero matrix elements: $\langle
e0|V_1|gk\rangle$, $\langle f0|V_2|ek\rangle$, and $\langle
ek|V_1|gk_1k_2\rangle$.

First we prove that $\langle e0|R(z)|f0\rangle=0$. From (\ref{21})
we obtain:
\begin{equation}\label{22}
\left\langle {e0} \right|R_{ef} (z)\left| {f0} \right\rangle  =
\sum\limits_k {} \frac{{\left\langle {e0} \right|V_1 \left| {gk}
\right\rangle \left\langle {gk} \right|R(z)\left| {f0}
\right\rangle }} {{z - \omega _k }}
\end{equation}

\begin{equation}\label{23}
\left\langle {gk} \right|R(z)\left| {f0} \right\rangle  =
\left\langle {gk} \right|V\frac{Q} {{z - H_0 }}R(z)\left| {f0}
\right\rangle
\end{equation}

Since $\langle gk|V|Q=0$ the right hand side of (\ref{23}) is
zero. Therefore from (\ref{22}) $\langle e0|R(z)|f0\rangle=0$ and
from (\ref{A3}) we obtain for $G_{ee}(z)$ , $G_{ff}(z)$ the
expressions (\ref{14}).

\subsection{Calculation of $
R_{ee} (z) = \left\langle {e0} \right|R(z)\left| {e0}
\right\rangle $ }

Applying (\ref{21}) we obtain:

\begin{equation}\label{Ree}
R_{ee} (z) = \left\langle {e0} \right|V\left| {e0} \right\rangle +
\left\langle {e0} \right|V\frac{Q} {{z - H_0 }}R(z)\left| {e0}
\right\rangle
\end{equation}

The first term in the right hand side of (\ref{Ree}) is equal to
zero, $\langle e0|V|e0\rangle=0$. In addition, there is the only
matrix element which transfer the state $|e0\rangle$ to the
continuum states $|g,k\rangle$ of the projector $Q$, namely,
$\langle e,0|V_1|g,k\rangle$. Therefore,

\begin{equation}\label{Ree1}
R_{ee}(z)= \sum\limits_k {} \frac{{\left\langle {e0} \right|V_1
\left| {gk} \right\rangle \left\langle {gk} \right|R(z)\left| {e0}
\right\rangle }} {{z - E_g  - \omega _k }}
\end{equation}

The calculation of $\left\langle {gk} \right|R(z)\left| {e0}
\right\rangle $ is straightforward. From (\ref{21}) we obtain:

\begin{equation}\label{Ree2}
\left\langle {gk} \right|R(z)\left| {e0} \right\rangle  =
\left\langle {gk} \right|V_1 \left| {e0} \right\rangle  +
\left\langle {gk} \right|V\frac{Q} {{z - H_0 }}R(z)\left| {e0}
\right\rangle
\end{equation}

The second term in the right hand side of (\ref{Ree2}) is zero
because the interaction $V$ transfer the states $|gk\rangle$ to
the state $|e0\rangle$ which is orthogonal to the projector $Q$.
Therefore, for (\ref{Ree1}) we finally obtain:

\begin{equation}\label{Ree3}
\left\langle {e0} \right|R (z)\left| {e0} \right\rangle  =
\sum\limits_k {} \frac{{\left| {\left\langle {e0} \right|V_1
\left| {gk} \right\rangle } \right|^2 }} {{z - E_g  - \omega _k }}
\end{equation}

Applying here the Sokhotski formula

\begin{equation}\label{Sok}
\mathop {\lim }\limits_{\eta  \to 0^ +  } \frac{1} {{E - \omega
\pm i\eta }} =  \mp i\pi \delta (E - \omega ) + Pv\left( {\frac{1}
{{E - \omega }}} \right)
\end{equation}

we obtain from (\ref{Ree3})
\begin{equation}\label{Ree4}
 \langle e,0|R(E\pm i\eta|e,0\rangle=\Delta_1(E)\mp i\frac{\Gamma_1(E)}{2}
\end{equation}
where

\begin{equation}\label{Delta1}
\Delta _1 \left( E \right) = Pv\sum\limits_k {} \frac{{\left|
{\left\langle {e0} \right|V_1 \left| {gk} \right\rangle }
\right|^2 }} {{E - E_g  - \omega _k }}
\end{equation}

\begin{equation}\label{Gam1}
\Gamma _1 (E) = 2\pi \sum\limits_{k } {} \left| {\left\langle {e0}
\right|V_1 \left| {gk} \right\rangle } \right|^2 \delta (E - E_g
- \omega _k )
\end{equation}

Finally, for $U_{ee}(E)$ we obtain

\begin{equation}\label{Uee}
U_{ee} (E) = \frac{1} {{2\pi }}\frac{{\Gamma _1^{} (E)}} {{\left(
{E - E_e  - \Delta _1 (E)} \right)^2  + \frac{{\Gamma _1^2 (E)}}
{4}}}
\end{equation}

The main nontrivial conclusion that follows from these expressions
is that the decay of the second transmon level, if excited, does
not depend at all on its interaction with the third level. This
result is general in nature: the decay of the third level is not
influenced by higher transmon levels. Therefore, we can draw a
general conclusion: the population dynamics of any excited level
of a transmon depends only on its interaction with the lower-lying
levels.This issue is briefly addressed in Appendix B.

\subsection{Calculation of $
R_{ff} (z) = \left\langle {f0} \right|R(z)\left| {f0}
\right\rangle $ }

Applying (\ref{21}) we obtain the following sequence of coupled
equations:

\begin{equation}\label{24}
\begin{gathered}
R_{ff} (z) = \left\langle {f0} \right|V\frac{Q} {{z - H_0
}}R(z)\left| {f0} \right\rangle \hfill\\ = \sum\limits_k {}
\frac{{\left\langle {f0} \right|V_2 \left| {ek} \right\rangle
\left\langle {ek} \right|R(z)\left| {f0} \right\rangle }} {{z -
E_e  - \omega _k }}
\end{gathered}
\end{equation}
\begin{equation}\label{25}
\begin{gathered}
  \left\langle {ek} \right|R(z)\left| {f0} \right\rangle  = \left\langle {ek} \right|V_2 \left| {f0} \right\rangle  + \left\langle {ek} \right|V\frac{Q}
{{z - H_0 }}R(z)\left| {f0} \right\rangle  \hfill \\
   = \left\langle {ek} \right|V_2 \left| {f0} \right\rangle  + \sum\limits_{k_1 ,k_2 } {} \frac{{\left\langle {ek} \right|V_1 \left| {gk_1 k_2 } \right\rangle \left\langle {gk_1 k_2 } \right|R(z)\left| {f0} \right\rangle }}
{{z - E_g  - \omega _{k_1 }  - \omega _{k_2 } }} \hfill \\
\end{gathered}
\end{equation}
\begin{equation}\label{26}
\begin{gathered}
\left\langle {gk_1 k_2 } \right|R(z)\left| {f0} \right\rangle  =
\left\langle {gk_1 k_2 } \right|V\frac{Q} {{z - H_0 }}R(z)\left|
{f0} \right\rangle \hfill\\  = \sum\limits_k {}
\frac{{\left\langle {gk_1 k_2 } \right|V_1 \left| {ek}
\right\rangle \left\langle {ek} \right|R(z)\left| {f0}
\right\rangle }} {{z - E_e  - \omega _k }}
\end{gathered}
\end{equation}
Substituting (\ref{26}) into (\ref{25}) yields:
\begin{equation}\label{27}
\begin{gathered}
\left\langle {ek} \right|R(z)\left| {f0} \right\rangle  =
\left\langle {ek} \right|V_2 \left| {f0} \right\rangle\hfill\\  +
\sum\limits_{k_1 ,k_2 } {} \sum\limits_{k'} {} \frac{{\left\langle
{ek} \right|V_1 \left| {gk_1 k_2 } \right\rangle \left\langle
{gk_1 k_2 } \right|V_1 \left| {ek'} \right\rangle \left\langle
{ek'} \right|R(z)\left| {f0} \right\rangle }} {{\left( {z - E_g  -
\omega _{k_1 }  - \omega _{k_2 } } \right)\left( {z - E_e  -
\omega _{k'} } \right)}}
\end{gathered}
\end{equation}

In the sum over $k^{'}$ in (\ref{27}) we leave only the term with
$k^{'}=k$.
\begin{equation}\label{28}
\begin{gathered}
  \left\langle {ek} \right|R(z)\left| {f0} \right\rangle  = \left\langle {ek} \right|V_2 \left| {f0} \right\rangle  \hfill \\
   + \sum\limits_{k_1 ,k_2 } {} \frac{{\left\langle {ek} \right|V_1 \left| {gk_1 k_2 } \right\rangle \left\langle {gk_1 k_2 } \right|V_1 \left| {ek} \right\rangle \left\langle {ek} \right|R(z)\left| {f0} \right\rangle }}
{{\left( {z - E_g  - \omega _{k_1 }  - \omega _{k_2 } } \right)\left( {z - E_e  - \omega _k } \right)}} \hfill \\
\end{gathered}
\end{equation}
The matrix element $\langle ek|R(z)|f0\rangle $ is readily
obtained from (\ref{28}):

\begin{equation}\label{29}
\left\langle {ek} \right|R(z)\left| {f0} \right\rangle  =
\frac{{\left\langle {ek} \right|V_2 \left| {f0} \right\rangle }}
{{1 - \sum\limits_{k_1 ,k_2 } {} \frac{{\left| {\left\langle {ek}
\right|V_1 \left| {gk_1 k_2 } \right\rangle } \right|^2 }}
{{\left( {z - E_g  - \omega _{k_1 }  - \omega _{k_2 } }
\right)\left( {z - E_e  - \omega _k } \right)}}}}
\end{equation}
The procedure which leads from (\ref{27}) to (\ref{29}) is
equivalent to the summing in series (\ref{Rz}) only the resonant
terms $(z-E_e-\omega_k)^n$, where $n$ is integer. Substituting
(\ref{29}) into (\ref{24}) yields:

\begin{equation}\label{30}
R_{ff} (z) = \sum\limits_k {} \frac{{\left| {\left\langle {ek}
\right|V_2 \left| {f0} \right\rangle } \right|^2 }} {{z - E_e  -
\omega _k  - \sum\limits_{k_1 ,k_2 } {} \frac{{\left|
{\left\langle {ek} \right|V_1 \left| {gk_1 k_2 } \right\rangle }
\right|^2 }} {{\left( {z - E_g  - \omega _{k_1 }  - \omega _{k_2 }
} \right)}}}}
\end{equation}

We note that the interaction $V_1$ between the second level
$|e\rangle$ and the ground state $|g\rangle$ influences the decay
of the third level $|f\rangle$. Below we show that the quantity
$R_{ff}(z)$ defines the frequency shift $ \Delta _2 (E,\eta )$ and
the width ${{\Gamma _2 (E,\eta )}}$ of the transmon qubit which
appear in (\ref{20}) .

\subsection{Calculation of $R_{ff}(E\pm i\eta)$ (\ref{18})}

Here we calculate the frequency shift $ \Delta _2 (E,\eta )$ and
the width ${{\Gamma _2 (E,\eta )}}$ of transmon.

From (\ref{30}) we have

\begin{equation}\label{32}
\begin{gathered}
R_{ff} (E \pm i\eta )\hfill\\ = \sum\limits_k {} \frac{{\left|
{\left\langle {ek} \right|V_2 \left| {f0} \right\rangle }
\right|^2 }} {{E \pm i\eta  - E_e  - \omega _k  - \sum\limits_{k_1
,k_2 } {} \frac{{\left| {\left\langle {ek} \right|V_1 \left| {gk_1
k_2 } \right\rangle } \right|^2 }} {{\left( {E \pm i\eta  - E_g  -
\omega _{k_1 }  - \omega _{k_2 } } \right)}}}}
\end{gathered}
\end{equation}
With the aid of Sokhotski formula (\ref{Sok}) we rewrite the
double sum in the denominator of (\ref{32}):

\begin{equation}\label{34}
\begin{gathered}
  \sum\limits_{k_1 ,k_2 } {} \frac{{\left| {\left\langle {ek} \right|V_1 \left| {gk_1 k_2 } \right\rangle } \right|^2 }}
{{\left( {E \pm i\eta  - E_g  - \omega _{k_1 }  - \omega _{k_2 } } \right)}} \hfill \\
   =  \mp i\pi \sum\limits_{k_1 ,k_2 } {} \left| {\left\langle {ek} \right|V_1 \left| {gk_1 k_2 } \right\rangle } \right|^2 \delta (E - E_g  - \omega _{k_1 }  - \omega _{k_2 } ) \hfill \\
   + \sum\limits_{k_1 ,k_2 } {} \frac{{\left| {\left\langle {ek} \right|V_1 \left| {gk_1 k_2 } \right\rangle } \right|^2 }}
{{E - E_g  - \omega _{k_1 }  - \omega _{k_2 } }} \equiv \Delta _1
\left( {E,\omega_k } \right) \mp i\frac{{\Gamma _1 (E,\omega_k )}}
{2} \hfill \\
\end{gathered}
\end{equation}
where
\begin{equation}\label{35}
\Delta _1 \left( {E,\omega_k } \right) = \sum\limits_{k_1 ,k_2 }
{} \frac{{\left| {\left\langle {ek} \right|V_1 \left| {gk_1 k_2 }
\right\rangle } \right|^2 }} {{E - E_g  - \omega _{k_1 }  - \omega
_{k_2 } }}
\end{equation}

\begin{equation}\label{36}
\Gamma _1 (E,\omega_k ) = 2\pi \sum\limits_{k_1 ,k_2 } {} \left|
{\left\langle {ek} \right|V_1 \left| {gk_1 k_2 } \right\rangle }
\right|^2 \delta (E - E_g  - \omega _{k_1 }  - \omega _{k_2 } )
\end{equation}

Therefore, for $R_{ff}(E\pm i\eta)$ we obtain

\begin{equation}\label{37}
\begin{gathered}
  R_{ff} (E \pm i\eta ) \hfill \\
   = \sum\limits_k {} \frac{{\left| {\left\langle {ek} \right|V_2 \left| {f0} \right\rangle } \right|^2 }}
{{E - E_e  - \omega _k  - \Delta _1 (E,\omega_k ) \pm i\left(
{\frac{{\Gamma _1 (E,\omega_k )}}
{2} + \eta } \right)}} \hfill \\
   = \sum\limits_k {} \frac{{\left| {\left\langle {ek} \right|V_2 \left| {f0} \right\rangle } \right|^2 \left( {\left( {E - E_e  - \omega _k  - \Delta _1 (E,\omega_k )} \right)} \right)}}
{{\left( {E - E_e  - \omega _k  - \Delta _1 (E,\omega_k )}
\right)^2 + \left( {\frac{{\Gamma _1 (E,\omega_k )}}
{2} + \eta } \right)^2 }} \hfill \\
   \mp i\frac{{\left( {\frac{{\Gamma _1 (E,\omega_k )}}
{2} + \eta } \right)}} {{\left( {E - E_e  - \omega _k  - \Delta _1
(E,\omega_k )} \right)^2  + \left( {\frac{{\Gamma _1 (E,\omega_k
)}}
{2} + \eta } \right)^2 }} \hfill \\
\end{gathered}
\end{equation}

Comparing (\ref{37}) with (\ref{18}) we obtain
\begin{equation}\label{38}
\begin{gathered}
\Delta _2 (E,\eta )\hfil\\ = \sum\limits_k {} \frac{{\left|
{\left\langle {ek} \right|V_2 \left| {f0} \right\rangle }
\right|^2 \left( {E - E_e  - \omega _k  - \Delta _1 (E,\omega_k )}
\right)}} {{\left( {E - E_e  - \omega _k  - \Delta _1 (E,\omega_k
)} \right)^2  + \left( {\frac{{\Gamma _1 (E,\omega_k )}} {2} +
\eta } \right)^2 }}
\end{gathered}
\end{equation}

\begin{equation}\label{39}
\begin{gathered}
\Gamma _2 (E,\eta )\hfill\\ = 2\sum\limits_k {} \frac{{\left|
{\left\langle {ek} \right|V_2 \left| {f0} \right\rangle }
\right|^2 \left( {\frac{{\Gamma _1 (E,\omega_k )}} {2} + \eta }
\right)}} {{\left( {E - E_e  - \omega _k  - \Delta _1 (E,\omega_k
)} \right)^2  + \left( {\frac{{\Gamma _1 (E,\omega_k )}} {2} +
\eta } \right)^2 }}
\end{gathered}
\end{equation}

\section{Transmon vs two-level system}

Here we compare the decay of transmon with that of a generic
two-level system. Due to the coupling $V_1$ between $|e\rangle$
and $|g\rangle$ the decay of transmon qubit differs significantly
from that of a two-level atom whose frequency shift and linewidth
are given by expressions (\ref{Delta1}) and (\ref{Gam1}). As seen
from equation (\ref{30}), the interaction between $|e\rangle$ and
$|g\rangle$ opens a fast two-photon decay channel via the matrix
element $\langle ek|V_1|gk_1,k_2\rangle$. In what follows, we
substantiate this statement using more general considerations
based directly on the Dyson equation for the resolvent.

\begin{equation}\label{Dy1}
G(z) = G_0 (z) + G_0 (z)VG(z)
\end{equation}

where $G_0^{ - 1} (z) = z - H_0$.

First, we consider the decay of a two-level system with an excited
state $|e\rangle$ and ground state $|g\rangle$. The interaction
between these states is given in (\ref{3}). In this case, we
restrict Hilbert space to the two vectors $|e,0\rangle$ and
single-photon states $|g,k\rangle$. The projector operators for
these states are $P=|e,0\rangle\langle e,0|$, $Q = \sum\limits_k {}
\left| {g,k} \right\rangle \left\langle {g,k} \right|$, satisfying
the completeness relation $P+Q=1$.

For the population $\langle e,0|G(z)|e,0\rangle$ and the
probability amplitude $\langle g,k|G(z)|e,0\rangle$ we obtain from
(\ref{Dy1}):

\begin{equation}\label{Dy2}
\begin{gathered}
  \left\langle {e0} \right|G(z)\left| {e0} \right\rangle  =
  \left\langle {e0} \right|G_0 (z)\left| {e0} \right\rangle
   + \left\langle {e0} \right|G_0 (z)V_1 G(z)\left| {e0} \right\rangle  \hfill \\
   = \frac{1}
{{z - E_e }} + \frac{1}
{{z - E_e }}\sum\limits_k {} \left\langle {e0} \right|V_1 \left| {gk}
\right\rangle \left\langle {gk} \right|G(z)\left| {e0} \right\rangle  \hfill \\
\end{gathered}
\end{equation}
 and
\begin{equation}\label{Dy3}
\left\langle {gk} \right|G(z)\left| {e0} \right\rangle  = \frac{1}
{{z - E_g  - \omega _k }}\left\langle {gk} \right|V_1\left| {e0}
\right\rangle \left\langle {e0} \right|G(z)\left| {e0}
\right\rangle
\end{equation}

From these two equations we get the following expression for the
population of the second level:

\begin{equation}\label{Dy4}
\left\langle {e0} \right|G(z)\left| {e0} \right\rangle  = \left(
{z - E_e  - \sum\limits_k {} \frac{{\left| {\left\langle {e0}
\right|V_1\left| {gk} \right\rangle } \right|^2 }} {{z - E_g  -
\omega _k }}} \right)^{ - 1}
\end{equation}

The sum over $k$ in (\ref{Dy4}) is precisely the quantity
(\ref{Ree3}).

Further, we consider the population of a third transmon level,
$\left\langle {f0} \right|G(z)\left| {f0} \right\rangle$. In this
case, the projectors $P$ and $Q$ are defined as in (\ref{P}) and
(\ref{Q}). From the Dyson equation (\ref{Dy1}) we obtain:

\begin{equation}\label{Dy5}
\begin{gathered}
  \left\langle {f0} \right|G(z)\left| {f0} \right\rangle  = \frac{1}
{{z - E_f }} \hfill \\
   + \frac{1}
{{z - E_f }}\sum\limits_k {} \left\langle {f0} \right|V_2 \left| {ek} \right\rangle \left\langle {ek} \right|G(z)\left| {f0} \right\rangle  \hfill \\
\end{gathered}
\end{equation}

It follows from (\ref{Dy5}) that the population $\left\langle {f0}
\right|G(z)\left| {f0} \right\rangle$ is described by the
superposition of transition amplitudes $\left\langle {ek}
\right|G(z)\left| {f0} \right\rangle$ with different photon
frequencies. This amplitude can, in turn, be expressed from the
Dyson equation as follows:

\begin{equation}\label{Dy6}
\begin{gathered}
  \left\langle {ek} \right|G(z)\left| {f0} \right\rangle  = \frac{1}
{{z - E_e  - \omega _k }}\left\langle {ek} \right|V_2 \left| {f0} \right\rangle \left\langle {f0} \right|G(z)\left| {f0} \right\rangle  \hfill \\
   + \frac{1}
{{z - E_e  - \omega _k }}\sum\limits_{p,q} {} \left\langle {ek} \right|V_1 \left| {gpq} \right\rangle \left\langle {gpq} \right|G(z)\left| {f0} \right\rangle  \hfill \\
\end{gathered}
\end{equation}

From (\ref{Dy5}) and (\ref{Dy6}) we obtain:

\begin{equation}\label{Dy7}
\begin{gathered}
  \left\langle {f0} \right|G(z)\left| {f0} \right\rangle  = \frac{1}
{{z - E_f }} \hfill \\
   + \frac{1}
{{z - E_f }}\sum\limits_k {} \frac{{\left| {\left\langle {f0}
\right|V_2 \left| {ek} \right\rangle } \right|^2 }}
{{z - E_e  - \omega _k }}\left\langle {f0} \right|G(z)\left| {f0} \right\rangle  \hfill \\
   + \frac{1}
{{z - E_f }}\sum\limits_{k,p,q} {} \frac{{\left\langle {f0}
\right|V_2 \left| {ek} \right\rangle \left\langle {ek} \right|V_1
\left| {gpq} \right\rangle }}
{{z - E_e  - \omega _k }}\left\langle {gpq} \right|G(z)\left| {f0} \right\rangle  \hfill \\
\end{gathered}
\end{equation}

The third term in (\ref{Dy7}) describes the influence of the
interaction between the second transmon level $|e\rangle$ and its
ground state $|g\rangle$. It is precisely the interaction
amplitude, $\langle ek|V_1|gpq\rangle$, which results in the
emission of two photons that survive in the long time limit. If we
disregard this term, we would obtain for $\langle
f0|G(z)|f0\rangle$ precisely the analog of the two-level equation
(\ref{Dy4}).

The calculations beyond the scope of this paper show that the
amplitude for a single-photon emission, $\langle
ek|G(z)|f0\rangle$, has two poles in the lower half of the complex
$z$-plane. The frequency integral of the Fourier transform of this
amplitude tends to zero as $t\rightarrow\infty$. On the other
hand, one of the poles of the two-photon amplitude, $\langle
gpq|G(z)|f0\rangle$, lies on the $z$-axis which  results in the
survival of the two photons as  $t\rightarrow\infty$.

In summary, we conclude that the interaction $V_1$ between
$|e\rangle$ and $|g\rangle$ opens a fast two-photon decay channel,
$\langle ek|V_1|gk_1k_2\rangle$, which broadens the second
transmon level $|e\rangle$. On the other hand, in a ladder type
three-level system the emission of two coherent photons is
impossible. Therefore, the two photons emitted via the interaction
between the second level and the first level are inherently
incoherent. This incoherence means it is impossible to determine
which photon is emitted by the $f\rightarrow e$ transition and
which by the $e\rightarrow g$ transition. This
indistinguishability leads to destructive interference between the
two paths.

\section{Resonances and widths of the third transmon level}

In this section, we find analytical expressions describing the
resonances and widths of a transmon's third level. As shown in
Section V, the coupling between the transmon's second level and
its ground state significantly influences the decay dynamics of
the third level. In order to compare the decay of the transmon
with that of a generic two-level system, we will formally turn off
the interaction $V_1$ between the second and first levels in
(\ref{20}).

Even though turning off $V_1$ is experimentally unfeasible, we use
this thought procedure here to compare the decay of the transmon
and that of a two-level atom. If the coupling between $|e\rangle$
and $|g\rangle$ were absent ($V_1=0$) the state $|e\rangle$ would
be stable. As seen from (\ref{35}) and (\ref{36}) in this case,
the quantities $\Delta_1(E,\omega_k)$, and $\Gamma_1(E,\omega_k)$,
are equal to zero. Therefore, for $V_1=0$ we obtain:

\begin{equation}\label{40}
\Delta _2 (E,\eta ) = \sum\limits_k {} \frac{{\left| {\left\langle
{ek} \right|V_2 \left| {f0} \right\rangle } \right|^2 \left( {E -
E_e  - \omega _k } \right)}} {{\left( {E - E_e  - \omega _k }
\right)^2  + \eta ^2 }}
\end{equation}

\begin{equation}\label{41}
\Gamma _2 (E,\eta ) = 2\sum\limits_k {} \frac{{\left|
{\left\langle {ek} \right|V_2 \left| {f0} \right\rangle }
\right|^2 \eta }} {{\left( {E - E_e  - \omega _k } \right)^2  +
\eta ^2 }}
\end{equation}

In the limit $\eta\rightarrow 0$ using the known expression
\begin{equation}\label{42}
\mathop {\lim }\limits_{\eta  \to 0_ +  } \frac{\eta } {{\left( {E
- E_e  - \omega } \right)^2  + \eta ^2 }} = \pi \delta (E - E_e  -
\omega )
\end{equation}
we obtain for (\ref{40}), (\ref{41})
\begin{equation}\label{43}
\Delta _2 (E,0) = \sum\limits_k {} \frac{{\left| {\left\langle
{ek} \right|V_2 \left| {f0} \right\rangle } \right|^2 }} {{\left(
{E - E_e  - \omega _k } \right)}}
\end{equation}

\begin{equation}\label{44}
\Gamma _2 (E,0) = 2\pi \sum\limits_k {} \left| {\left\langle {ek}
\right|V_2 \left| {f0} \right\rangle } \right|^2 \delta (E - E_e -
\omega _k )
\end{equation}

Therefore, in this case, for the function $U_{ff}(E)$ we obtain:

\begin{equation}\label{45}
U_{ff} (E) = \frac{1} {{2\pi }}\frac{{\Gamma _2^{} (E,0)}}
{{\left( {E - E_f  - \Delta _2 (E,0)} \right)^2  + \frac{{\Gamma
_2^2 (E,0)}} {4}}}
\end{equation}

With the substitution of $|f,0\rangle$, $|e,k\rangle$, $E_f$, and
$V_2$ with $|e,0\rangle$, $|g,k\rangle$, $E_e$, and $V_1$,
respectively, the expressions (\ref{43}), (\ref{44}) coincide
exactly with (\ref{Delta1}) and (\ref{Gam1}) describing the decay
of two-level atom. Therefore, the expression  (\ref{45}) exactly
matches (\ref{Uee}) describing the decay of excited state
$|f\rangle$ if the second state $|e\rangle$ were stable.

If $V_1\neq 0$ we may put $\eta=0$ directly in (\ref{20}):

\begin{equation}\label{46}
U_{ff} (E) = \frac{1} {{2\pi }}\frac{{\Gamma _2^{} (E)}} {{\left(
{E - E_f  - \Delta _2 (E)} \right)^2  + \frac{{\Gamma _2^2 (E)}}
{4}}}
\end{equation}
where
\begin{equation}\label{47}
\Delta _2 (E) = \sum\limits_k {} \frac{{\left| {\left\langle {ek}
\right|V_2 \left| {f0} \right\rangle } \right|^2 \left( {E - E_e -
\omega _k  - \Delta _1 (E,\omega_k )} \right)}} {{\left( {E - E_e
- \omega _k  - \Delta _1 (E,\omega_k )} \right)^2  + \frac{{\Gamma
_1^2 (E,\omega_k )}} {4}}}
\end{equation}

\begin{equation}\label{48}
\Gamma _2 (E) = 2\sum\limits_k {} \frac{{\left| {\left\langle {ek}
\right|V_2 \left| {f0} \right\rangle } \right|^2 \frac{{\Gamma _1
(E,\omega_k )}} {2}}} {{\left( {E - E_e  - \omega _k  - \Delta _1
(E,\omega_k )} \right)^2  + \frac{{\Gamma _1^2 (E,\omega_k )}}
{4}}}
\end{equation}

The quantities $\Delta_1(E,\omega_k)$, and $\Gamma_1(E,\omega_k)$
in (\ref{47}), (\ref{48}) describe the influence of the coupling
between $|e\rangle$ and $|g\rangle$ on the decay of the state
$|f\rangle$.

The resonance points of $U_{ff}(E)$ (\ref{46}) are defined by, in
general, the nonlinear equation
\begin{equation}\label{67}
E - E_f  - \Delta _2 (E) = 0
\end{equation}

In order to proceed the calculations we need only two matrix
elements:

\begin{equation}\label{49}
\begin{gathered}
  \left\langle {ek} \right|V_2 \left| {f0} \right\rangle  = g_2 (k) \hfill \\
  \left\langle {gk_1 k_2 } \right|V_1 \left| {ek} \right\rangle  = \frac{1}
{{\sqrt 2 }}\left( {g_1 (k_1 )\delta _{k_2 ,k}  + g_1 (k_2 )\delta _{k_1 ,k} } \right) \hfill \\
\end{gathered}
\end{equation}

Therefore, we obtain

\begin{equation}\label{50}
\Delta _2 (E) = \sum\limits_k {} \frac{{g_2^2 (k)\left( {E - E_e -
\omega _k  - \Delta _1 (E,\omega_k )} \right)}} {{\left( {E - E_e
- \omega _k  - \Delta _1 (E,\omega )} \right)^2  + \frac{{\Gamma
_1^2 (E,\omega_k )}} {4}}}
\end{equation}

\begin{equation}\label{51}
\Gamma _2 (E) = \sum\limits_k {} \frac{{g_2^2 (k)\Gamma _1
(E,\omega_k )}} {{\left( {E - E_e  - \omega _k  - \Delta _1
(E,\omega_k )} \right)^2  + \frac{{\Gamma _1^2 (E,\omega_k )}}
{4}}}
\end{equation}

where

\begin{equation}\label{52}
\Delta _1 \left( {E,\omega _k } \right) = \sum\limits_{k'} {}
\frac{{g_1^2 (k')}} {{E - \omega _{k'}  - \omega }} + \frac{{g_1^2
(k)}} {{2(E - 2\omega _k )}}
\end{equation}

\begin{equation}\label{53}
\begin{gathered}
\Gamma _1 (E,\omega _k ) = 2\pi \sum\limits_{k'} {} g_1^2
(k')\delta (E - \omega _k  - \omega _{k'} )\hfill\\ + \pi g_1^2
(k)\delta (E - 2\omega _k )
\end{gathered}
\end{equation}

Expression (\ref{50}) allows one to determine the resonance
frequency shift by solving the nonlinear equation (\ref{67}). The
number of roots of this equation depends on the strength of the
transmon's interaction with the electromagnetic field. Each
solution of this equation corresponds to a resonance width given
by formula (\ref{51}). Quantities (\ref{52}) and (\ref{53})
describe the contribution of the interaction between the
transmon's second level and its ground state. As follows from
(\ref{50}) and (\ref{51}), each of the quantities (\ref{52}) and
(\ref{53}) enters both expression (\ref{50}) and expression
(\ref{51}).

\section{Transformation to continuum modes}

We begin with discrete quantities $g_1(k),g_2(k)$ in (\ref{3}),
(\ref{4}) which are defined as follows:

\begin{equation}\label{54}
g_i (k) = \sqrt {\frac{{ \omega _k d_i^2 }} {{2\varepsilon _0\hbar
LA}}} ,\;\quad (i = 1,2)
\end{equation}

where $A$ is the effective cross section of a waveguide, $L$ its
length, $d_i$ has a dimension of a dipole moment which for
transmon is expressed in terms of Josephson and charging energies
\cite{Koch2007}. The quantities $g_i(k)$ in (\ref{54}) have a
dimension of frequency.

In the continuum limit sums over discrete quantities $k$ are
converted to integrals over continuous frequency $\omega$
according to the rule:

\begin{equation}\label{54a}
\sum\limits_k {}  \to \frac{1} {{\Delta \omega
}}\int\limits_0^\infty  {d\omega }
\end{equation}

where
\begin{equation}\label{55}
\Delta \omega  = \frac{{2\pi v_g }} {L},
\end{equation}

$v_g$ is the group velocity of the guided modes in a waveguide,
$L$ is the length of a cavity.

The mode spectrum becomes continuous as $L\rightarrow\infty$ and
$\Delta\omega\rightarrow 0$.  In this limit the transformation
from discrete creation and destruction operator $a_k^+, a_k$ to
their continuous counterparts $a^+(\omega), a(\omega)$ is realized
according to the rules $ a_k  \to (\Delta \omega )^{1/2} a(\omega
);{\kern 1pt} \quad a_k^ +   \to (\Delta \omega )^{1/2} a^ +
(\omega )$ \cite{Blow1990}. In continuous limit we must take
$L\rightarrow\infty$, so that in the final expressions the
dependence of $L$ must be absent.

For example, the continuum mode expression for $V_1$ (\ref{3}) is
as follows:
\begin{equation}\label{56}
V_1  = \int\limits_0^\infty  {d\omega } g_1 (\omega )\left(
{\left| g \right\rangle \left\langle e \right|a^ +  (\omega ) +
\left| e \right\rangle \left\langle g \right|a(\omega } \right)
\end{equation}

where
\begin{equation}\label{57}
g_1 (\omega ) = \sqrt {\frac{{ \omega d_1^2 }} {{4\pi \varepsilon
_0\hbar v_g A}}}
\end{equation}
has a dimension of $\sqrt{\omega}$. Similarly, $ \sum\limits_k
{g_1^2 (k)}  \to \int\limits_0^\infty  {d\omega } g_1^2 (\omega )
$ where $g_1(\omega)$ is defined in (\ref{57}).

Therefore, the last terms in (\ref{52}), (\ref{53}) disappear for
$L\rightarrow\infty$. Finally, for the quantities (\ref{50}),
(\ref{51}), (\ref{52}), (\ref{53}) we obtain their continuous-mode
counterparts.
\begin{equation}\label{58}
\Delta _2 (E) = \int\limits_0^\infty  {d\omega } \frac{{g_2^2
(\omega )\left( {E - E_e  - \omega  - \Delta _1 (E,\omega )}
\right)}} {{\left( {E - E_e  - \omega  - \Delta _1 (E,\omega )}
\right)^2  + \left( {\frac{{\Gamma _1 (E,\omega )}} {2}} \right)^2
}}
\end{equation}

\begin{equation}\label{59}
\Gamma _2 (E) = \int\limits_0^\infty  {d\omega } \frac{{g_2^2
(\omega )\Gamma _1 (E,\omega )}} {{\left( {E - E_e  - \omega  -
\Delta _1 (E,\omega )} \right)^2  + \left( {\frac{{\Gamma _1
(E,\omega )}} {2}} \right)^2 }}
\end{equation}

where
\begin{equation}\label{60}
\Delta _1 \left( {E,\omega } \right) = Pv\int\limits_0^\infty
{d\omega '} \frac{{g_1^2 (\omega ')}} {{E - \omega ' - \omega }}
\end{equation}

\begin{equation}\label{61}
\Gamma _1 (E,\omega ) = 2\pi g_1^2 (E - \omega )
\end{equation}

\section{The coupling between transmon and continuum modes.
Density of states}

We consider here a one-dimensional waveguide that supports photon
modes up to several GHz. The flat density of states (DOS)
corresponds to ideal infinite waveguide, which is never truth in
the real experiments. In fact, a real waveguide looks rather like
a broadband cavity with the low quality factor. In this limit, the
waveguide has no well-defined resonant modes, but rather a
continuous spectrum of propagating modes. It is reasonable to
assume that a qubit embedded in such a waveguide effectively
interacts only with the frequencies which lie within a full width
at half maximum (FWHM) bandwidth, $2\gamma$ relative to the qubit
excitation frequency $\Omega$. This can be expressed as the
Lorentzian density of states

\begin{equation}\label{Lor}
\rho_L (\omega ) = \frac{1} {\gamma }\frac{1} {{1 + \frac{{(\omega
- \Omega )^2 }} {{\gamma ^2 }}}}
\end{equation}

where the quantity $\gamma$ represents the rate with which the
radiation deposited in the cavity is dissipated  due to losses
through the walls, open ends, etc.

For convenience of the subsequent simulations, instead of
(\ref{Lor}) we choose here the Gaussian density of states

\begin{equation}\label{GDOS}
\rho _G (\omega ) = \frac{2} {{\sqrt {\pi } \delta }}\exp \left( {
- \frac{{(\omega  - \Omega )^2 }} {{\delta ^2 }}} \right)
\end{equation}

which is assumed to be fitted with (\ref{Lor}) by equating their
FWHMs.

The Gaussian DOS is typically associated  with  structured
environments \cite{Br2016, Long2006}. For our numerical
simulations, we employ a Gaussian DOS due to its mathematical
convenience.

We define the collective coupling between transmon levels and
continuum modes $\sqrt{\Lambda_i}$ as in \cite{Zeb2022}:

\begin{equation}\label{63}
\Lambda _i  = \int_0^\infty  {d\omega } g_i^2 (\omega );\;i=1,2
\end{equation}

where indices $1,2$ correspond to the levels $|e\rangle$ and
$|f\rangle$, respectively.

Next, we introduce the distribution function $P_i(\omega)$ defined
as

\begin{equation}\label{62}
g_i^2 (\omega ) = \Lambda _i P_i(\omega );\;i = 1,2
\end{equation}

where

\begin{equation}\label{64}
\int_0^\infty  {P_i(\omega )d\omega }  = 1
\end{equation}
which is consistent with the definition (\ref{63}).

In fact, $P_i(\omega)$ is the density of states that describes the
distribution of the interactions $g_i(\omega)$ within the
continuum.

 Below we assume the coupling $g_i(\omega)$ is localized in a
FWHN region of the function $P_i(\omega)$ \cite{Zeb2022} around
the transition energies of the transmon qubit. For the
distribution $P_i(\omega)$ we take a Gaussian functions.

\begin{equation}\label{65}
P_1 (\omega ) = \frac{2} {{\delta \sqrt \pi  }}e^{ -
\frac{{(\omega  - (E_e  - E_g ))^2 }} {{\delta ^2 }}}  = \frac{2}
{{\delta \sqrt \pi  }}e^{ - \frac{{(\omega  - E_e )^2 }} {{\delta
^2 }}}
\end{equation}
where we set $E_g=0$,
\begin{equation}\label{P2}
P_2 (\omega ) = \frac{2} {{\delta \sqrt \pi  }}e^{ -
\frac{{(\omega  - (E_f  - E_e ))^2 }} {{\delta ^2 }}}  = \frac{2}
{{\delta \sqrt \pi  }}e^{ - \frac{{(\omega  - E_e  + \Delta E_e
)^2 }} {{\delta ^2 }}}
\end{equation}

where $ E_f  - E_e  = E_e ^\prime   = E_e  - \Delta E_e$.

In (\ref{65}), (\ref{P2}) $\delta$ is the  width of continuum
modes which effectively interact with a transmon. For coupling we
thus obtain

\begin{equation}\label{66}
g_1^2 (\omega ) = \frac{{2\Lambda _1 }} {{\delta \sqrt \pi  }}e^{
- \frac{{(\omega  - E_e )^2 }} {{\delta ^2 }}}
\end{equation}

\begin{equation}\label{g2}
    g_2^2 (\omega ) =  \frac{2\Lambda_2}
{{\delta \sqrt \pi  }}e^{ - \frac{{(\omega  - E_e  + \Delta E_e
)^2 }} {{\delta ^2 }}}
\end{equation}

From (\ref{66}) and (\ref{g2}), it follows that the interaction
strength between transmon qubit  and continuous modes can be
described by the dimensionless parameter $\Lambda_i/\delta^2$.
Therefore, in our case, the interaction strength is measured
relative to mode bandwidth $\delta$ which effectively interacts
with the transmon.

It should be noted that this definition should not be confused
with the generally accepted definition in waveguide QED where the
interaction strength between qubit and the continuum is defined in
terms of the ratio $g/\Omega$ where $\Omega$ is the excitation
frequency of a qubit. In our case, the coupling strength $g_{1,2}$
is assumed to be much less than $E_f$ or $E_e$. Therefore, from
conventional point of view we remain in a weak coupling regime.

In this paper, we consider the interaction weak if
$\Lambda_i/\delta^2\ll 1$ ( $\sqrt{\Lambda_i}\ll\delta$) and strong if
$\Lambda_i/\delta^2\geq 1$. ( $\sqrt{\Lambda_i}\geq\delta$).

A weak interaction describes the Markovian regime, in which the
resonance width essentially remains constant within the bandwidth
$\delta$. In this case, one can state that near the qubit
transition the DOS is flat. If $\sqrt{\Lambda}\ll\delta$, the
relaxation of the continuum, $1/\delta$ is much faster than the
qubit relaxation $1/\sqrt{\Lambda}$. Therefore, the future
interactions of qubit do not overlap with previous interactions as
the continuum erases the things much faster.

The non-Markovian regime corresponds to the case
$\sqrt{\Lambda}\geq\delta$, in which the resonance width depends
primarily on the energy within the bandwidth. In this case, we can
say that the qubit interacts with the continuum faster than the
continuum can erase its past. It is somewhat analogous to the
resonator case, though not identical. In this way, the Gaussian
DOS allows us to smoothly explore the transition between Markovian
and non-Markovian regimes.

\subsection{Weak coupling}
For weak interaction we expect $\Delta_2(E)\ll E_f$, therefore we
may approximate $\Delta_2(E)$ and $\Gamma_2(E)$ with
$\Delta_2(E_f)$ and $\Gamma_2(E_f)$, correspondingly. For this
case we obtain from (\ref{58}), (\ref{59}), (\ref{60}),
(\ref{61}):

\begin{equation}\label{68}
\Delta _2 (E_f ) = \frac{{2\Lambda _2 }} {{\delta \sqrt \pi
}}\int\limits_0^\infty  {d\omega } \frac{{e^{ - \frac{{(\omega  -
E_e +\Delta E_e)^2 }} {{\delta ^2 }}} \left( {E_e  - \omega-\Delta
_1 (E_f,\omega ) } \right)}} {{\left( {E_e  - \omega } \right)^2 +
4\pi \frac{{\Lambda _1^2 }} {{\delta ^2 }}e^{ - \frac{{2(\omega -
E_e )^2 }} {{\delta ^2 }}} }}
\end{equation}

\begin{equation}\label{69}
\Gamma _2 (E_f ) = \frac{{8\Lambda _2 \Lambda _1 }} {{\delta ^2
}}\int\limits_0^\infty  {d\omega } \frac{{e^{ - \frac{{(\omega -
E_e  + \Delta E_e )^2 }} {{\delta ^2 }}} e^{ - \frac{{(\omega  -
E_e )^2 }} {{\delta ^2 }}} }} {{\left( {E_e  - \omega } \right)^2
+ \frac{{2\pi \Lambda _1^2 }} {{\delta ^2 }}e^{ - \frac{{2(\omega
- E_e )^2 }} {{\delta ^2 }}} }}
\end{equation}

\begin{equation}\label{70}
\Delta _1 \left( {E_f ,\omega } \right) = \frac{{2\Lambda _1 }}
{{\delta \sqrt \pi  }}Pv\int\limits_0^\infty  {d\omega '}
\frac{{e^{ - \frac{{(\omega ' - E_e )^2 }} {{\delta ^2 }}} }}
{{E_f  - \omega ' - \omega }}
\end{equation}

\begin{equation}\label{71}
\Gamma _1 (E_f ,\omega ) = 2\pi \frac{{2\Lambda _1 }} {{\delta
\sqrt \pi  }}e^{ - \frac{{(\omega  - E_e )^2 }} {{\delta ^2 }}}
\end{equation}

\subsection{Strong coupling}

If we switch off the interaction between the second level of
transmon $|e\rangle$ and the ground state $|g\rangle$ ($V_1=0$) we
obtain from (\ref{43}), (\ref{44}):

\begin{equation}\label{72}
\Delta _2 \left( {E,0} \right) = \frac{{2\Lambda _2 }} {{\delta
\sqrt \pi  }}Pv\int\limits_0^\infty  {d\omega } \frac{{e^{ -
\frac{{(\omega  - E_e  + \Delta E_e )^2 }} {{\delta ^2 }}} }} {{E
- E_e  - \omega }}
\end{equation}

\begin{equation}\label{73}
\Gamma _2 (E,0) = 2\pi \frac{{2\Lambda _2 }} {{\delta \sqrt \pi
}}e^{ - \frac{{(E - 2E_e  + \Delta E_e )^2 }} {{\delta ^2 }}}
\end{equation}

The equations (\ref{72}), (\ref{73}) describe the frequency shift
and the width of  the third level $|f\rangle$,  assuming the
second level $|e\rangle$ were stable. In fact, this decay is
equivalent to the decay of a two-level system.

If the second level $|e\rangle$ interacts with the ground state
$|g\rangle$, ($V_1\neq 0$), then we use the equations (\ref{58}),
(\ref{59}) to obtain:

\begin{equation}\label{76}
\begin{gathered}
 \Delta _2 (E)\hfill\\ = \frac{{2\Lambda _2 }} {{\sqrt \pi  \delta
}}\int\limits_0^\infty  {d\omega } \frac{{e^{ - \frac{{(\omega  -
E_e  + \Delta E_e )^2 }} {{\delta ^2 }}} \left( {E - E_e  - \omega
- \Delta _1 (E,\omega )} \right)}} {{\left( {E - E_e  - \omega  -
\Delta _1 (E,\omega )} \right)^2  + \left( {\frac{{\Gamma _1
(E,\omega )}} {2}} \right)^2 }}
\end{gathered}
\end{equation}

\begin{equation}\label{77}
\begin{gathered}
\Gamma _2 (E)\hfill\\ = \frac{{8\Lambda _1 \Lambda _2 }} {{\delta
^2 }}\int\limits_0^\infty  {d\omega } \frac{{e^{ - \frac{{(\omega
- E_e  + \Delta E_e )^2 }} {{\delta ^2 }}} e^{ - \frac{{(E - E_e -
\omega )^2 }} {{\delta ^2 }}} }} {{\left( {E - E_e  - \omega  -
\Delta _1 (E,\omega )} \right)^2  + \left( {\frac{{\Gamma _1
(E,\omega )}} {2}} \right)^2 }}
\end{gathered}
\end{equation}
where
\begin{equation}\label{78}
\Delta _1 \left( {E,\omega } \right) = \frac{{2\Lambda _1 }}
{{\delta \sqrt \pi  }}Pv\int\limits_0^\infty  {d\omega '}
\frac{{e^{ - \frac{{(\omega ' - E_e )^2 }} {{\delta ^2 }}} }} {{E
- \omega ' - \omega }}
\end{equation}

\begin{equation}\label{79}
\Gamma _1 (E,\omega ) = 2\pi \frac{{2\Lambda _1 }} {{\delta \sqrt
\pi  }}e^{ - \frac{{(E - E_e  - \omega )^2 }} {{\delta ^2 }}}
\end{equation}

The quantities $\Delta_2(E)$ and $\Gamma_2(E)$ are the energy
dependent frequency shift and the width of the third level
$|f\rangle$ respectively, taking into account the values
$\Delta_1(E,\omega)$ and $\Gamma_1(E,\omega)$ which describe the
interaction of the second level $|e\rangle$  with the ground
state.

\section{Numerical results and discussions}

In this section, we calculate the resonances of the function
$U_{ff}(E)$ and their widths. For our calculations we use the
dimensionless quantities $E/\delta\equiv y$, $E_e/\delta\equiv a$,
$E_f/\delta\equiv b$. For the calculation of the spectra we take
the typical values for transmon levels: $E_e=2\pi\times 5$ GHz,
$E_f=2\pi\times 9.85$ GHz \cite{Sult2025}. Therefore, the
frequency anharmonicity $\Delta E_e=2\pi\times 150$ MHz. The
choice of the continuum width, $\delta$ is somewhat arbitrary. We
take $\delta=2\pi\times100$ MHz. Therefore, $a=50$ , $b=98.5$. We
study the dependence of the decay spectra on the dimensionless
parameters $L_1=\Lambda_1/\delta^2$,
$L_2=\frac{3}{2}\Lambda_1/\delta^2$ which characterizes the
coupling strength between the qubit and the continuum.

\begin{figure}
  \includegraphics[width=8 cm]{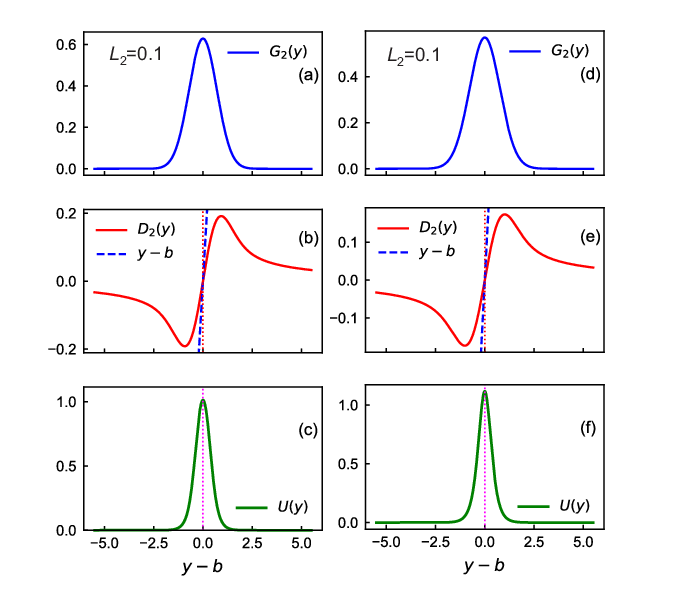}\\
  \caption{Transmon decay parameters for weak coupling, $L_2=0.1$. The
  left column describes the decay of the third level when
   the second level is stable ($V_1=0$). The influence of the interaction between
   $|e\rangle$ and $|g\rangle$ is shown in the right column where
   $L_1=\frac{2}{3}L_2$. The plots in left column are calculated from
   (\ref{72}), (\ref{73}), (\ref{45}), while those in right column are calculated from
  (\ref{76}), (\ref{77}), (\ref{46}).}\label{Fig2}
\end{figure}

\begin{figure}
  \includegraphics[width=8 cm]{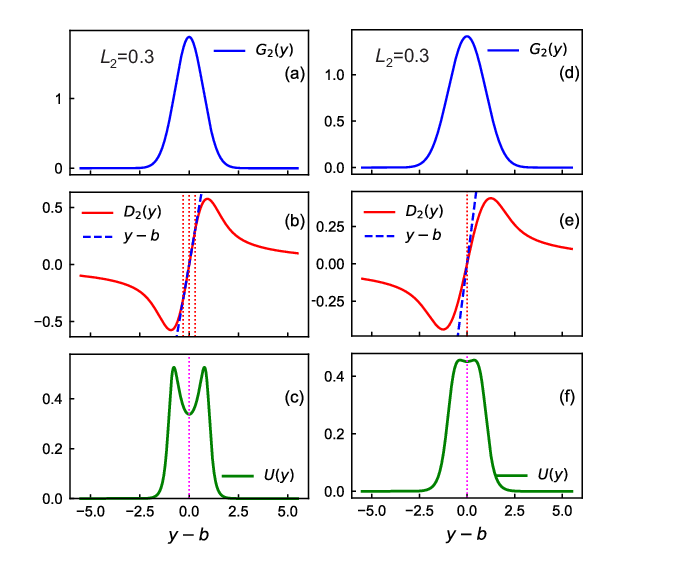}\\
  \caption{Transmon decay parameters for relatively weak coupling, $L_2=0.3$. The
  left column describes the decay of the third level when
   the second level is stable ($V_1=0$). The influence of the interaction between
   $|e\rangle$ and $|g\rangle$ is shown in the right column where
   $L_1=\frac{2}{3}L_2$. The plots in left column are calculated from
   (\ref{72}), (\ref{73}), (\ref{45}), while those in right column are calculated from
  (\ref{76}), (\ref{77}), (\ref{46}).}\label{Fig3}
\end{figure}

\begin{figure}
  \includegraphics[width=8 cm]{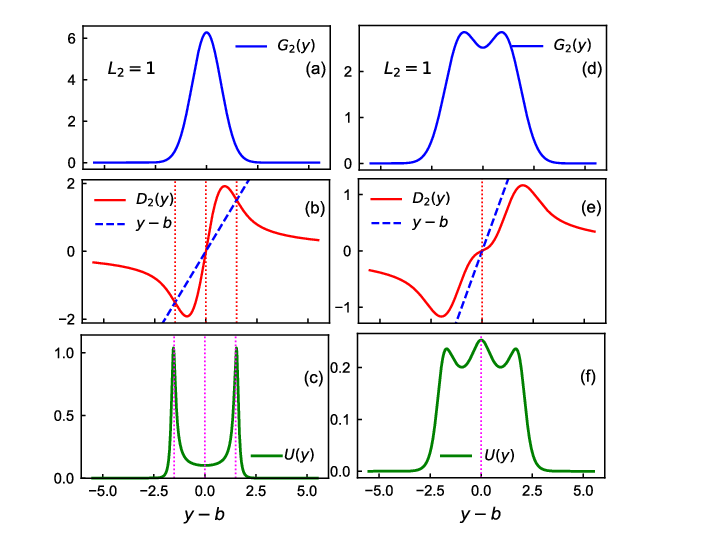}\\
  \caption{Transmon decay parameters for strong coupling, $L_2=1$. The
  left column describes the decay of the third level when
   the second level is stable ($V_1=0$). The influence of the interaction between
   $|e\rangle$ and $|g\rangle$ is shown in the right column where
   $L_1=\frac{2}{3}L_2$. The plots in left column are calculated from
   (\ref{72}), (\ref{73}), (\ref{45}), while those in right column are calculated from
  (\ref{76}), (\ref{77}), (\ref{46}).}\label{Fig5}
\end{figure}

\begin{figure}
  \includegraphics[width=8 cm]{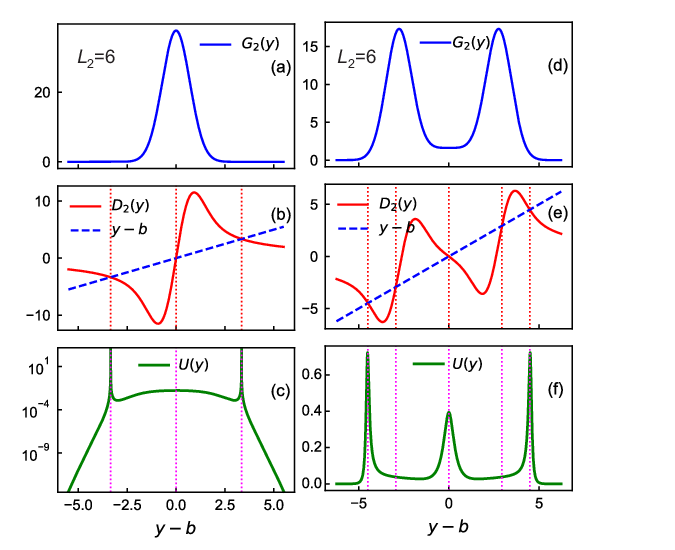}\\
  \caption{Transmon decay parameters for coupling, $L_2=6$. The
  left column describes the decay of the third level when
   the second level is stable ($V_1=0$). The influence of the interaction between
   $|e\rangle$ and $|g\rangle$ is shown in the right column where
   $L_1=\frac{2}{3}L_2$. The plots in left column are calculated from
   (\ref{72}), (\ref{73}), (\ref{45}), while those in right column are calculated from
  (\ref{76}), (\ref{77}), (\ref{46}).}\label{Fig4}
\end{figure}

In what follows we numerically study  the influence of the
interaction between the second level $|e\rangle$ and ground state
$|g\rangle$ on the decay of the third level $|f\rangle$. We
systematically compare the decay  of the third level when $V_1=0$
with that when $V_1\neq 0$. If $V_1=0$ the second level is stable
and the decay of the third level is identical with the decay of
the excited state of a two-level atom. The results of our numerics
are shown in the Fig.\ref{Fig2}, Fig.\ref{Fig3}, Fig.\ref{Fig5},
Fig.\ref{Fig4}.

In these figures, the width, $\Gamma_2(E)$, the frequency shift,
$\Delta_2(E)$, and spectral function, $U_{ff}(E)$ are given in the
dimensionless units, $G_2(y)=\Gamma_2(E)/\delta$,
$D_2(y)=\Delta_2(E)/\delta$, $U(y)=U_{ff}(E)\delta$, respectively.
The figures are arranged in a three-plot two columns. The left
column shows the results for $V_1=0$ when the second level is
stable. For this case, the plots have been numerically calculated
from equations (\ref{72}), (\ref{73}), (\ref{45}). For comparison
the results for $V_1\neq 0$ are shown in the right column, where
$L_2$ is the same as in the left column and $L_1=\frac{2}{3}L_2$.
In this case, the equations (\ref{76}), (\ref{77}), (\ref{46})
have been used for numerical calculations. Every column, which
consists of three plots, shows the dependence on the dimensionless
frequency detuning, $E/\delta-E_f/\delta\equiv y-b$. The
dependence of the width, $\Gamma_2(E)/\delta$ is shown in the
first plot. The frequency shift $\Delta_2(E)/\delta$ is shown in
the second plot. And finally, the dimensionless spectral function,
$U_2(E)\delta$ is shown in the third plot.

In general, the width $\Gamma_2(E)>0$ and  rapidly tends to zero
as $E\rightarrow\infty$, while the frequency shift $\Delta_2(E)$
resembles the shape of dispersive curve which changes the sign in
the vicinity of $E_f$ \cite{Kohen2004}.

The distinctive feature of our plots is that the frequency shift
$\Delta_2(E=E_f)=0$ both for $V_1=0$ and $V_1\neq 0$. It is the
consequence of the Gaussian density of states $P(\omega)$
(\ref{65}), (\ref{P2}), which allows for the shifting the lower
bound in the frequency integrals (\ref{72}), (\ref{76}) to
$-\infty$.

The crossover between weak and strong interaction occurs when
there appear  two additional intersections on the frequency shift
curve. Every root of the equation (\ref{67}) results in a peak in
the function $U_{ff}(E)$, (see Fig.\ref{Fig3}, left column).

\begin{figure}
  \includegraphics[width=8 cm]{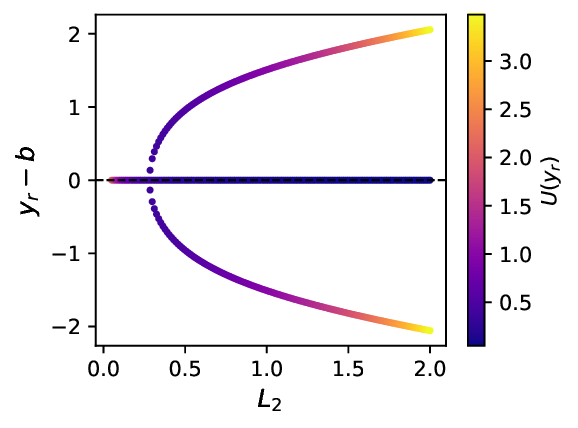}\\
  \caption{Dependence of the intersection points $y_r$ of straight line
  $y-b$ with the dispersive curve $D_2(y)$ on the coupling parameter $L_2$ for $V_1=0$.
  The colour bar shows the value $U(y_r)$ in the intersection point $y_r$.}\label{Fig6}
\end{figure}

As $L_2$ exceeds the crossover point two peaks move apart from
each other as is shown in Fig.\ref{Fig6}. In this regime the
Wigner-Weisscopf approximation is not valid, the width
$\Gamma_2(E)$ is reduced quickly with the detuning $E-E_f$.
Therefore, the spectral function $U_{ff}(E)$ is no longer a
Lorentzian.

As $L_2$ becomes larger the peaks become narrower and their
heights are also increased. It is a signature of the formation of
long lived quasistable states in the continuum. The beat note of
the Fourier transform of the two narrow peaks gives rise to the
slowly decaying Rabi oscillations in the time domain.

The dependence of intersection points on the value of $L_2$ for
$V_1=0$ is shown in Fig.\ref{Fig6}. The crossover is at
$L_2\approx0.28$. For $0 < L_2 < 0.28$, the line $y-b$ intersects
the curve $D_2(y)$ at the only point $y-b=0$. This interval
corresponds to the straight segment on the $x$-axis in
Fig.\ref{Fig6}. For $L_2 > 0.28$, in addition to the intersection
at zero, two more intersection points appear, symmetrically
located with respect to zero. Thus, each value of $L_2$ for $L_2
> 0.28$ corresponds to three intersection points, i.e., three
resonances. The color scale represents the magnitudes of these
resonances, which correspond to the peaks in the lower panel of
the left column in Fig.\ref{Fig2}(c)- Fig.\ref{Fig4}(c). Note that
the resonance at the point $y-b=0$ does not produce a peak
because, at this point, the resonance widths in Fig.\ref{Fig2}(a)-
Fig.\ref{Fig4}(a) are relatively large.

 As an example, we consider the bottom panel in the left column in
 Fig.\ref{Fig4}. The calculated positions of the resonances are
 $E_{r1}=95.15\times\delta$, $E_{r2}=101.84\times\delta$ with the equal
 heights $U_{ff,max}=1230/\delta$. Therefore, for $\delta=2\pi\times100$ MHz we obtain:
$E_{r1}=2\pi\times9.515$ GHz, $E_{r2}=2\pi\times10.184$ GHz, The
form of resonance curves is not Lorentzian therefore, we cannot
estimate the  widths of these resonances from (\ref{45}),
$U_{ff,max}=2/(\pi \Gamma_2(E_r,0))$. From numerical simulations
we have found  full width at half maximum (FWHM)
$\Gamma_2(E_r,0)=2\pi\times24$ kHz.

Thus, in this case, the frequency of Rabi oscillations
$(E_{r2}-E_{r1})/2=2\pi\times0.33$ GHz with  the oscillation
period $T_R=3\times 10^{-9}$ s and the decay rate
$\tau=1/(\Gamma_2(E,0))=6.6\times 10^{-6}$ s. As $T_R\ll\tau$ the
Rabi oscillations are well resolved.

\begin{figure}
  \includegraphics[width=8 cm]{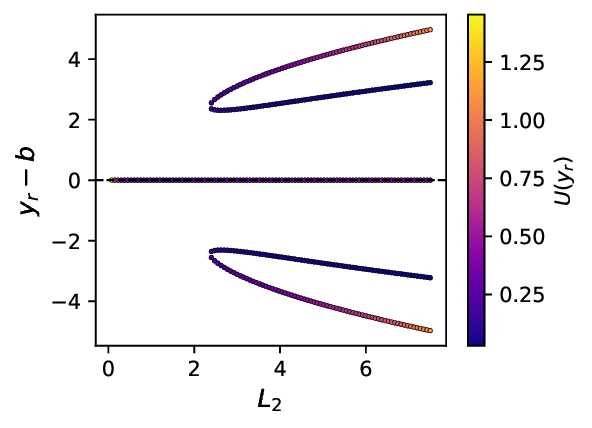}\\
  \caption{Dependence of the intersection points $y_r$ of straight line
  $y-b$ with the dispersive curve $D_2(y)$ on the coupling parameter
  $L_2$ for $V_1\neq 0$.
  The colour bar shows the value $U(y_r)$ in the intersection point $y_r$.}\label{Fig7}
\end{figure}

The case $V_1\neq 0$ is shown in the right columns in the
Fig.\ref{Fig2}, Fig.\ref{Fig3}, Fig.\ref{Fig5}, Fig.\ref{Fig4}.
Comparing these plots with those from the left column we see that
the interaction between the second level and the ground state
significantly alters the picture of two-level decay shown in the
left columns. First, the peaks in $U_{ff}(E)$ is not necessarily
associated with the roots of equation (\ref{67}) (see
Fig.\ref{Fig5} e, f). The intersections begin at relatively large
values of $L_2$ (see Fig.\ref{Fig7}). Second, not all roots of
equation (\ref{67}) give rise to the peaks in $U_{ff}(E)$. For
example, there are five intersections in Fig.\ref{Fig4}e, however,
there are only three peaks in $U_{ff}(E)$ which are associated
with relatively low values of $\Gamma_2(E)$ in Fig.\ref{Fig4}d.
Third, the heights of the peaks are greatly reduced as compared to
two-level decay.

The dependence of intersection points on the value of $L_2$ for
the case $V_1\neq0$ is shown in Fig.\ref{Fig7}. Here, the
crossover between weak and strong coupling also begins at
$L_2\approx 0.28$. However, unlike in Fig.\ref{Fig6}, the
intersection of the line $y-b$ and the curve $D_2(y)$ starts
approximately at $L_2\approx2$. For $0<L_2<2$, the intersection
occurs only at zero. In this region, the width of the peak
$G_2(y)$ begins to depend significantly on energy. The line $y-b$
closely approaches the curve $D_2(y)$, which manifests as
additional small peaks in the dependence U(y) (see the right
column of Fig.\ref{Fig5}). For $L_2>2$, additional resonances
arise, corresponding to the intersection points of the line $y-b$
and the curve $D_2(y)$. In Fig.\ref{Fig4}(e), for example, we see
four additional intersection points. However, not all of them lead
to visible peaks on the energy curve $U(y)$. This is due to the
strong dependence of the width $G_2(y)$ on energy. The farther an
intersection point is from zero, the smaller the width $G_2(y)$
corresponding to that point. Therefore, for these distant points
the resonances are more pronounced (see Fig.\ref{Fig4}(f)).

For comparison with the two-level case, we consider the bottom
panel in the right column in
 Fig.\ref{Fig4}. The calculated positions of the resonances are
 $E_{r1}=94.02\times\delta$, $E_{r2}=98.5\times\delta$, $E_{r3}=102.98\times\delta$
 with the
 heights $U_{ff,max,1}=0.723/\delta$, $U_{ff,max,2}=0.394/\delta$,
  $U_{ff,max,3}=0.723/\delta$. Therefore, for $\delta=2\pi\times100$ MHz we obtain:
$E_{r1}=2\pi\times9.402$ GHz, $E_{r2}=2\pi\times9.85$ GHz
$E_{r3}=2\pi\times10.248$ GHz. The calculated FWHM of resonances,
$\Gamma_2(E_{r1})=\Gamma(E_{r3})=2\pi\times 22$ MHz,
$\Gamma_2(E_{r2})=2\pi\times 68$ MHz. Thus, in this case, the
frequency of Rabi oscillations $(E_{r3}-E_{r1})/2=2\pi\times0.423$
GHz with  the oscillation period $T_R=2.36\times 10^{-9}$ s and
the decay rate $\tau=1/(\Gamma_2(E_{r1}))=7.2\times 10^{-9}$ s. In
this case, $T_R\simeq\tau$, therefore, in this case, the Rabi
oscillations are not well resolved.

Therefore, for a two-level decay scheme the strong coupling gives
rise to the formation within a continuum of the quasi stable long
lived states which in turn give rise to a weak damping coherent
Rabi oscillations of the probability amplitude (Fig.\ref{Fig4})c.
However, the interaction between second level and ground state
completely destroys this coherence (Fig.\ref{Fig4})f.

In a ladder type three level system the emission of two coherent
photons is impossible. Therefore, the two photons emitted via the
interaction between the second level and the first level are
inherently incoherent. This incoherence means it is impossible to
determine which photon is emitted by the $f\rightarrow e$
transition and which by the $e\rightarrow g$ transition. This
indistinguishability leads to destructive interference between the
two paths.

This fact is reflected in the central peak in Fig.\ref{Fig4}(f).
It is this peak that gives rise to the sequential Rabi
oscillations. These three peaks are already visible in
Fig.\ref{Fig5})(f), which are due to the strong non Markovian
dependence of the level width on energy, as seen in
Fig.\ref{Fig5}(d) and Fig.\ref{Fig4}(d).

\section{Conclusion}

In this paper, we present a theoretical framework for calculating
the decay properties of the third transmon level coupled to a
broadband one-dimensional continuum of cavity modes. We compare in
detail the decay of the transmon with that of a generic two-level
system. We identify two distinct dynamical regimes, differentiated
by the ratio of the qubit's coupling strength to the continuum
bandwidth. When this ratio is much less than one, the system
exhibits a Markovian regime where the resonance width is
practically independent of energy within the continuum band. As
the ratio increases, the system transitions to a non-Markovian
regime where the resonance width becomes strongly
energy-dependent. We show that the coupling between the second
transmon level and the transmon ground state significantly
influences the decay of the third level. In the two-level decay
scenario, strong coupling leads to the formation of quasi-stable,
long-lived states within the continuum, which give rise to weakly
damped Rabi oscillations of the probability amplitude. However,
the interaction between the second level and the ground state
completely destroys this coherence. The destruction of coherence
can be attributed to the indistinguishability between two emitted
photons: one from the $f\rightarrow e$ transition and the other
from the $e\rightarrow g$ transition. This indistinguishability
leads to destructive interference between the two emission paths.

Although our focus here is specifically on a transmon
implementation, the derivation we present can be applied to any
generic ladder-type system. We argue that the results of this work
can be extended and are applicable to research on artificial atoms
in open quantum systems.

\begin{acknowledgments}

Ya.S.G. acknowledges the fruitful discussions with A. N. Sultanov.
The work is supported by the Ministry of Science and Higher
Education of Russian Federation under the project FSUN-2026-0004.
O.V.K. acknowledges the financial support from the Foundation
for the Advancement of Theoretical Physics and Mathematics
``BASIS''.

\end{acknowledgments}

\textbf{Data availability}

The data that support the findings of this article are available
from the authors upon reasonable request.

\appendix

\section{Application of projection operators technique
for the calculation of
the matrix elements of the resolvent}

The matrix elements of Green function $G(z)$ between the states of
the projector $P$ (\ref{P}) are defined as follows
\cite{Kohen2004}:

\begin{equation}\label{A1}
PG(z)P = \left( {\begin{array}{*{20}c}
   {G_{ee} (z)} & {G_{ef} (z)}  \\
   {G_{fe} (z)} & {G_{ff} (z)}  \\
\end{array} } \right)
\end{equation}
 where, for example $G_{ff}(z)=\langle f0|G(z)|f0\rangle$. This
 matrix element allows for the calculation of the probability that
 the system being initially in the state $|f0\rangle$ will remain
 in the state $|f0\rangle$ at some later time $t$. The matrix
 (\ref{A1}) is the inverse of the matrix \cite{Kohen2004}

\begin{equation}\label{A2}
\left( {\begin{array}{*{20}c}
   {z - E_e  - R_{ee} (z)} & {R_{ef} (z)}  \\
   {R_{fe} (z)} & {z - E_f  - R_{ff} (z)}  \\
\end{array} } \right)
\end{equation}

where the quantities  $R_{ff}=\langle f0|R(z)|f0\rangle$, etc. are
the matrix elements of the shift operator $R(z)$, the calculation
of which  for the interaction $V=V_1+V_2$ is the main goal of our
study.

Therefore

\begin{equation}\label{A3}
\begin{gathered}
  G_{ee} (z) = \frac{{z - E_f  - R_{ff} (z)}}
{{D(z)}};G_{ef} (z) = \frac{{R_{ef} (z)}}
{{D(z)}} \hfill \\
  G_{fe} (z) = \frac{{R_{fe} (z)}}
{{D(z)}};G_{ff} (z) = \frac{{z - E_e  - R_{ee} (z)}}
{{D(z)}} \hfill \\
\end{gathered}
\end{equation}
where
\begin{equation}\label{A4}
\begin{gathered}
D(z) = \left( {z - E_e  - R_{ee} (z)} \right)\left( {z - E_f -
R_{ff} (z)} \right)\hfill\\
- R_{ef} (z)R_{fe} (z)\\
\end{gathered}
\end{equation}

We prove in the main text that for the interaction $V=V_1+V_2$
where $V_1$ and $V_2$ are given in (\ref{3}) and (\ref{4}), the
off-diagonal elements of the shift operator $R_{ef}(z)$,
$R_{fe}(z)$ are exactly equal to zero. Therefore,
$G_{ef}(z)=G_{fe}(z)=0$ and we obtain

\begin{equation}\label{A5}
\begin{gathered}
  G_{ee} (z) = \frac{1}
{{\left( {z - E_e  - R_{ee} (z)} \right)}};\hfill \\
  G_{ff} (z) = \frac{1}
{{\left( {z - E_f  - R_{ff} (z)} \right)}} \hfill \\
\end{gathered}
\end{equation}

\section {The influence of the higher transmon levels on the decay of the third level}

The transmon has a ladder structure in which each level interacts
only with its neighboring levels. Consider a four-level transmon
with the fourth level denoted as $|h\rangle$. In this case,
Hamiltonian (\ref{1}) should be supplemented by
$E_h|h\rangle\langle h|$ and the additional interaction term:
\begin{equation}\label{B1}
V_3  = \sum\limits_k {} g_3 (k)\left( {\left| f \right\rangle
\left\langle h \right|a_k ^ +   + \left| h \right\rangle
\left\langle f \right|a_k } \right)
\end{equation}

The projector $P$ becomes

\begin{equation}\label{B2}
P = \left| {e0} \right\rangle \left\langle {e0} \right| + \left|
{f0} \right\rangle \left\langle {f0} \right| + \left| {h0}
\right\rangle \left\langle {h0} \right|
\end{equation}

The resolvent matrix is thus the inverse of a $3\times 3$ matrix,
which is similar to (\ref{A2}):

\begin{equation}\label{B3}
\left( {\begin{array}{*{20}c}
   {z - E_h  - R_{hh} } & {R_{hf} } & {R_{he} }  \\
   {R_{fh} } & {z - E_f  - R_{ff} } & {R_{fe} }  \\
   {R_{eh} } & {R_{ef} } & {z - E_e  - R_{ee} }  \\
\end{array} } \right)
\end{equation}
where $R_{ij}=\langle i0|R(z)|j0\rangle$ with $i,j=g, e, f , h$.

 The important point here is that if the third level of
transmon $|f\rangle$ is initially excited, then we can restrict
Hilbert space to three photon states (\ref{8}). Thus, in this
case, the projector $Q$ (\ref{Q}) remains the same. It then
follows that all off-diagonal elements in the matrix (\ref{B3})
are zero.

As an example we calculate the matrix element $R_{hf}=\langle
h0|R(z)|f0\rangle$. With the aid of (\ref{21}) we obtain

\begin{equation}\label{B4}
\left\langle {h0} \right|R(z)\left| {f0} \right\rangle  =
\left\langle {h0} \right|V\left| {f0} \right\rangle  +
\left\langle {h0} \right|V\frac{Q} {{z - H_0 }}R(z)\left| {f0}
\right\rangle
\end{equation}

As $\left\langle {h0} \right|V = \sum\limits_k {} g_3
(k)\left\langle {fk} \right|$, and the photon state $\left\langle
{fk} \right|$ is outside of our truncated Hilbert space (\ref{8}),
two matrix elements in (\ref{B4}) are equal to zero. Therefore,
$R_{hf}=\langle h0|R(z)|f0\rangle=0$. Similarly, it can be shown
that other off-diagonal elements in (\ref{B3}) also vanish, and
the quantity $R_{ff}(z)$ does not depend on $V_3$. Therefore,the
resolvent $G_{ff}(z)$, as expressed in (\ref{A5}), does not depend
on $V_3$ either. Thus, we can conclude that the decay of the third
transmon level does not depend on its interaction with the higher
level $|h\rangle$.

\end{document}